# Fano-Qubits for Quantum Devices with Enhanced Isolation and Bandwidth

Deepanshu Trivedi[1], Leonid Belostotski[2], Arjuna Madanayake[1], and Alex Krasnok[1,*]

[1]Department of Electrical and Computer Engineering, Florida International University, Miami, Florida 33174, USA

[2]Department of Electrical and Software Engineering, University of Calgary, Canada

[*]Email: akrasnok@fiu.edu

**Abstract**

*Magneto-optical isolators and circulators have been widely used to safeguard quantum devices from reflections and noise in the readout stage. However, these devices have limited bandwidth, low tunability, are bulky, and suffer from high losses, making them incompatible with planar technologies such as circuit QED. To address these limitations, we propose a new approach to quantum non-reciprocity that utilizes the intrinsic nonlinearity of qubits and broken spatial symmetry. We show that a circuit containing Lorentz-type qubits can be transformed into Fano-type qubits with an asymmetric spectral response, resulting in a significant improvement in isolation (up to 40 dB) and a twofold increase in spectral bandwidth (up to 200 MHz). Our analysis is based on realistic circuit parameters, validated by existing experimental results, and supported by rigorous quantum simulations. This approach could enable the development of compact, high-performance, and planar-compatible non-reciprocal quantum devices with potential applications in quantum computing, communication, and sensing.*

As classical systems, such as computers or sensors, approach their performance limits, quantum computers are emerging as exciting replacements with the potential to surpass classical computational systems' limitations[1]–[3]. Superconducting quantum circuits (circuit QEDs) operate at ultra-low temperatures to eliminate thermal noise and fluctuations due to the fragility of quantum states. However, to connect these quantum devices to measurement instruments operating at room temperature, a microwave isolator with unidirectional signal propagation is required[4], [5]. Isolators rely on nonreciprocity, where electromagnetic radiation propagates asymmetrically between two points. Ferrite materials' magneto-optical (Faraday) effect is typically used to achieve this effect at microwaves[6]–[8], but these devices are lossy, narrowband, barely tunable, bulky, and incompatible with planar circuit QED. While other approaches based on two-dimensional magnetic materials, topological isolators/semimetals, and cold atoms have also been extensively investigated recently (see, e.g.,[9]), they are still limited in many aspects and unsuitable for superconducting circuit QED systems. Another approach based on time-modulated resonators[10] requires an external energy input, precise control over the modulation phase, and connection to the modulation signal generators, making it prone to thermal noise. Therefore, there is a pressing need to develop a microwave isolator that is lossless, wide-band, tunable, compact, and compatible with superconducting circuit QED systems.

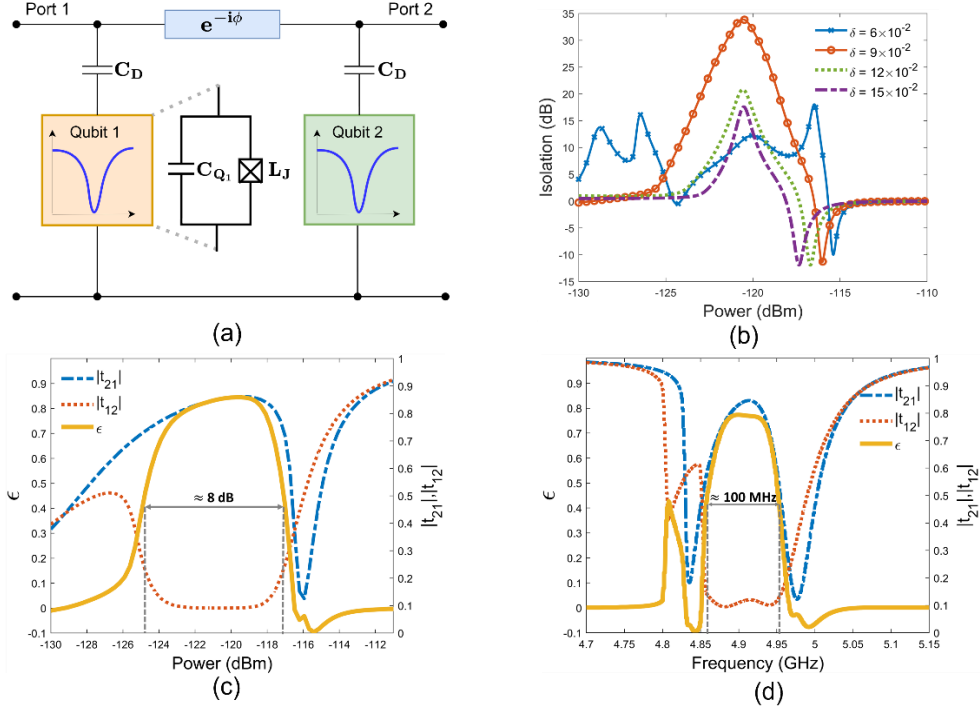

*Figure 1.* (a) Lumped-element schematic of a Lorentz-type qubit built up from Josephson junctions, separated by a distance $d$ (phase retardation $\phi$). (b) Isolation as a function of driving power for different detuning coefficients $\delta$. (c,d) Nonreciprocity evaluated as a function of (c) power, (d) frequency, for the forward transmission $|t_{21}|$ (blue), backward transmission $|t_{12}|$ (orange), and diode efficiency $\epsilon$ (yellow).

In this work, we revisit a recently proposed approach to quantum nonreciprocity that employs the intrinsic quantum nonlinearity of two-level systems, or qubits, combined with broken spatial symmetry[11]. The nonreciprocal behavior arises from the saturable scattering property of a single qubit, which reflects at low powers and transmits at high powers, as predicted in Ref.[12], experimentally demonstrated in Ref.[13], and theoretically elucidated in Ref.[9]. To enhance the efficacy of achieving nonreciprocal effects, a two-qubit system should be designed to support superradiant and subradiant states in a strong coupling regime. This design ensures asymmetric coupling to waves from opposing directions, leading to disparate levels of transparency for the qubits when subjected to excitation from distinct directions. Despite their potential advantages, the implementation of nonreciprocal designs in practical quantum systems remains a challenging task due to their limited bandwidth and impracticality, as discussed in earlier studies[14]. It is worth emphasizing that these issues represent fundamental restrictions that must be overcome to realize the full potential of nonreciprocal elements in quantum systems. To overcome this challenge, we propose modifying the system by converting a circuit with Lorentz-type qubits into one with Fano-type qubits with an asymmetric spectral response. This modification will substantially enhance the feasibility of achieving nonreciprocal effects in realistic quantum systems. Our analysis, based on realistic circuit parameters validated by existing experimental results and supported by rigorous quantum simulations, demonstrates that this modification results in a significant improvement in isolation (up to 40 dB) and a twofold increase in spectral bandwidth (up to 200 MHz). The modified system retains a slight improvement in the power bandwidth as well, making it a substantial enhancement over the previous approach.

First, we examine a two-qubit isolator depicted in Fig. 1(a), which exhibits Lorentz-type behavior. The qubits have a symmetrical structure, exhibiting robust quantum nonlinearity and marginally detuned resonant frequencies. Two coupling capacitors connect both qubits to the same transmission line, while a phase delay element ensures sufficient phase retardation between scattered waves. The circuit parameters used in simulations are corroborated by earlier experimental findings and thorough quantum simulations executed in QuCAT[15], as illustrated in Fig. 2.

We assume the system under investigation to be a circuit QED (cQED) system with Josephson junctions (JJs)[16]–[19], operating at temperatures below the superconducting critical temperatures of about 20 mK. This minimizes thermal noise in the frequency range of interest (5-10 GHz). The incorporation of JJs in the circuit simplifies the system to a two-level system with states $|0\rangle$ and $|1\rangle$, while anharmonicity separates all other states from the fundamental transition ( $\hbar\omega_{01} = \hbar(\omega_1 - \omega_0)$ ). The current through the junction, denoted by $I = I_c \sin(\phi)$, has a critical current $I_c$ associated with the Josephson energy $E_J$ through the relation $I_c = (2e/\hbar)E_J = (2\pi/\Phi_0)E_J$. In this case, $\Phi_0 = 2\pi\hbar/(2e)$ represents the magnetic flux quantum, $e$ signifies the electron charge, and $\phi = \phi_L - \phi_R$ is the Josephson phase of the two superconductors. The time variation of the Josephson phase yields a voltage drop $V$ across the superconductors, given by $d\phi(t)/dt = 2\pi V/\Phi_0$. A JJ has an intrinsic inductance ($L_J$), defined as $L_J = \Phi_0/(2\pi I_c \cos\phi) = L_{J0}/\sqrt{1-(I/I_c)^2}$, where $L_{J0}$ refers to the inductance at zero current. The $L_J$ dependence on $I$ implies that the JJ acts as a nonlinear inductor[20]. By adding a shunt capacitance ($C_Q$) across the JJ such that $E_J/E_C \gg 1$, a transmon qubit is realized. In this study, we select circuit parameters that operate the circuits in the transmon regime.

The qubits possess marginally detuned fundamental transition frequencies due to a variation in the shunt capacitance $C_{Q_2}$ by a factor of $\delta$ [$C_{Q_1} = C_{Q_2}(1+\delta)$], resulting in a resonant frequency shift of $|\eta| = e^2/2\hbar C_Q$ [4]. This design displays nonlinear transmission behavior based on the driven port. The qubits are separated by a distance $d$, defined as the inter-qubit spacing, given by $d = \lambda(1-\delta/\pi)/2$, where $\lambda$ is the wavelength of the operating frequency, and $\delta$ is the optimal detuning coefficient. The distance $d$ determines the drive phase $\phi$ between qubits. We simulate the circuit using Keysight Path Wave Advanced Design System (ADS) electronic design automation software, obtaining the steady-state solution through harmonic balance simulation. Fig. 1(a) features an insert representing the circuit of a single qubit, and we assume a realistic finite value of internal resistance ($R_J$) associated with the JJ. The simulation employs characteristic parameter values of ($C_{Q_1} = 68.67$ fF, $C_{Q_2} = 63$ fF, $C_D = 60$ fF, $F = 4.9$ GHz, $P_{in} = -123$ dBm, $\delta = 9 \times 10^{-2}$, and $R_J = 0.5\Omega$), which are typical for such circuits[19]. The critical current ($I_c$) for the transmon qubit is approximately 40 nA, corresponding to an $L_{J0}$ value of 8 nH. The design features $E_J/E_C = 67$, a value within the typical range for this parameter in transmon qubits.

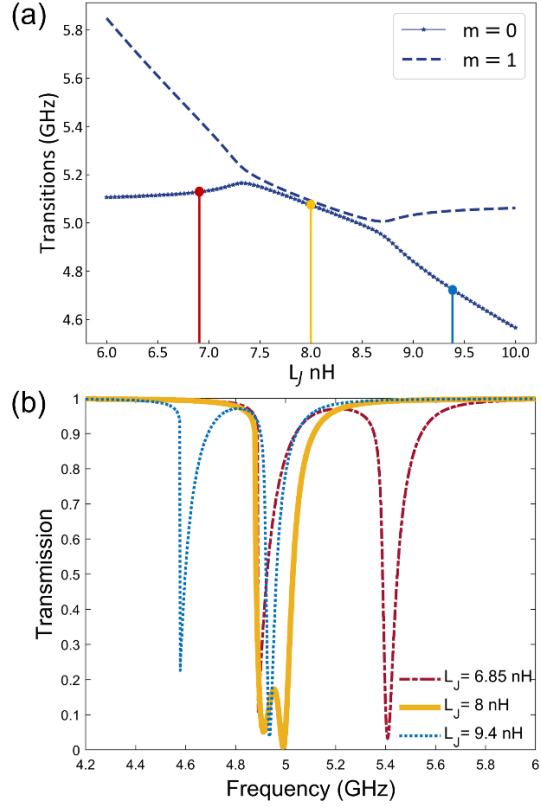

*Figure 2.* *(a) First two transition frequencies (eigenvalues) as a function of Josephson inductance ($L_J$) calculated using QuCAT. (b) Transmission of the isolator based on Lorentz qubits for three different values of Josephson Inductance ($L_J$) obtained using ADS.*

In Fig. 1(b), we illustrate the forward-to-backward transmission ratio (isolation), which is a key parameter in characterizing the system's isolation performance. This ratio is shown as a function of input wave power for a range of detuning coefficients. The isolation figure of merit, a crucial metric for system performance, exhibits a noticeable increase when the input power surpasses a certain threshold value. This enhancement leads to optimal performance within a distinct power range. The most advantageous outcome occurs at $\delta = 9 \times 10^{-2}$.

When investigating the relationship between power and isolation figure of merit by tuning the frequency within the $\pm 500$ MHz range, it is crucial to recognize that the detuning coefficient may deviate by up to $2\%$ to achieve the best performance in terms of isolation figure of merit. Such variations can cause detuning of the transition frequency and must be taken into account when interpreting experimental results.

Isolator efficiency, another essential characteristic for describing the isolation effect, is dependent on the forward and backward transmission parameters as follows:

$$\epsilon \equiv |t_{21}| \cdot \frac{|t_{21}| - |t_{12}|}{|t_{21}| + |t_{12}|}. \tag{1}$$

Figs. 1(c),(d) display the isolation efficiency as a function of both frequency and input driving power, revealing the system's power (frequency) bandwidth of approximately 9 dB (100 MHz). These findings are consistent with previously reported experimental results found in Ref.[13].

The Hamiltonian of the circuit based on the Lorentz-like qubits illustrated in Fig. 1(a) is given by: $\mathcal{H}_T = \mathcal{H}_1 + \mathcal{H}_2 + \mathcal{H}_{C_{D1}} + \mathcal{H}_{C_{D2}}$, where $\mathcal{H}_1(\mathcal{H}_2)$ is the Hamiltonian for transmon 1 (transmon 2) and $\mathcal{H}_{C_D}$ is the Hamiltonian of the coupling capacitance $C_D$. The Hamiltonian of the transmons is given by

$$\mathcal{H}_i(\Phi_i, Q_i) = \frac{\hat{Q}_i^2}{2C_{Q_i}} - E_{Q_i} \cos(\frac{\hat{\Phi}_i}{\phi_0}), \tag{2}$$

where i = 1,2,.... and $\hat{Q}_i$ is the total charge on the $i$-th junction, $\hat{\Phi}_i$ is the generalized magnetic flux across the $i$-th junction, and $\phi_0$ is the magnetic flux quantum given by $\phi_0 = \hbar/2e$, $E_{Q_1}(E_{Q_2})$ is the Josephson energy for transmon 1 (transmon 2) and $C_{Q_1}$ ($C_{Q_2}$) is the shunt capacitance for transmon 1 (transmon 2). The Hamiltonian of the coupling capacitances is $\mathcal{H}_{C_{Di}} = \hat{Q}_{C_{Di}}^2/2C_D$, where $\hat{Q}_{C_{Di}}$ is the charge on the coupling capacitance $C_D$.

We utilize the powerful and reliable Python Quantum Circuit Analyzer Tool (QuCAT) to efficiently compute the eigenvalues of the system's Hamiltonian. QuCAT has been extensively validated and is a versatile quantum simulation platform that accurately predicts system behavior, optimizes quantum circuits, and efficiently handles large-scale simulations[5], [15]. The Hamiltonian is generated by applying the Hamiltonian method to the circuit and is expressed using various parameters, such as circuit normal modes, Taylor expansion for junction non-linearities, the number of excitations for each mode, and unspecified component values, including $L_J$ in this study. As illustrated in Fig. 2(a), the first two transition frequencies (eigenvalues) are plotted as a function of Josephson inductance ($L_J$). We conduct a parametric sweep for the first qubit with varying values of $L_J$, while keeping the second qubit fixed at $L_J = 8$ nH. Fig. 2(a) reveals the presence of an avoided crossing near the symmetry point, $L_J = 8$ nH. At the other two points, specifically $L_J = 6.85$ nH and $L_J = 9.4$ nH, the eigenvalues of the Hamiltonian are farther apart, indicating a diminished coupling between the two modes. We compare these results with those obtained using ADS for three distinct values of $L_J$, as displayed in Fig. 2(b). The ADS simulations yield a splitting of the peak for $L_J = 8$ nH, signifying a strong coupling within the system. For $L_J = 6.85$ nH and $L_J = 9.4$ nH, we observe two separate peaks, which correspond to the two distinct transition frequencies. Consequently, the agreement between these results validates the accuracy of the ADS simulations.

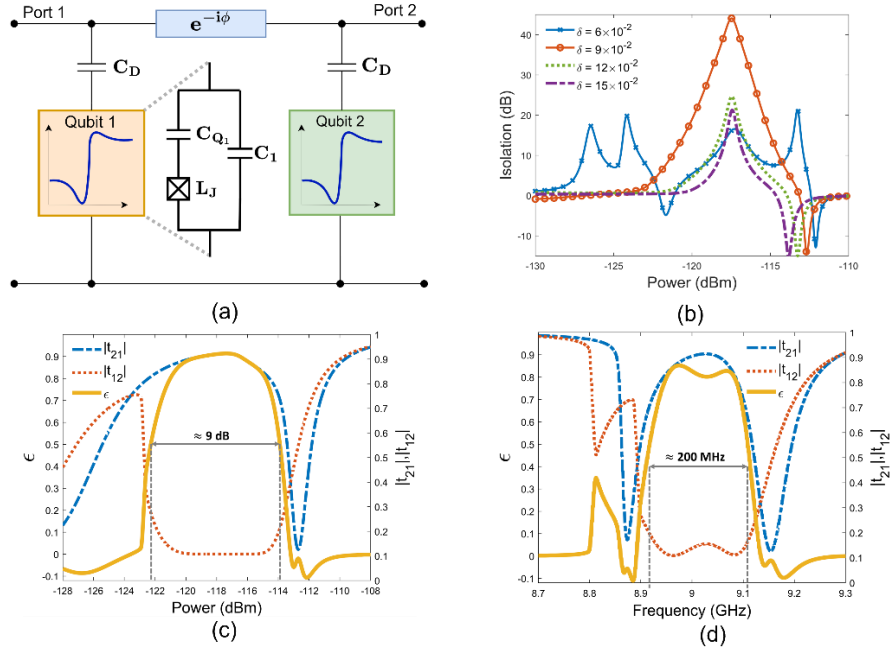

*Figure 3. (a) Lumped-element schematic of a Fano-type qubit built up from Josephson junctions, separated by a distance $d$ (phase retardation $\phi$ ). (b) The figure of merit as a function of driving power for different detuning coefficients $\delta$. Nonreciprocity evaluated as a function of (c) Power, (d) Frequency, for the forward transmission $|t_{21}|$ (blue), backward transmission $|t_{12}|$ (orange), and diode efficiency $\epsilon$ (yellow).*

Next, we establish that it is feasible to transform Lorentz-type qubits into Fano-type qubits, characterized by an asymmetric spectral response, by adapting the circuit design depicted in Fig. 3(a). Fano resonators are distinguished by their sharp and asymmetric frequency response, which leads to a swift transition from low to high transmission as the input intensity surpasses a specific threshold. Fig. 3(b) portrays the system's isolation, or the forward-to-backward transmission ratio, as a function of the input power for various detuning coefficients. The simulation employs the subsequent characteristic parameter values: (

$C_{Q_1} = 68.67$ fF, $C_{Q_2} = 63$ fF, $C_D = 60$ fF, $C_1 = C_2 = 26$ fF, $L_J = 8$ nH, $F = 8.98$ GHz, $P_{in} = -120$ dBm, $\delta = 9 \times 10^{-2}$, $R_J = 0.5$ Ω, and $I_c = 40$ nA). At $\delta = 9 \times 10^{-2}$, the system achieves optimal isolation, exhibiting an enhancement of approximately 10 dB over the ideal conditions for the Lorentzian circuit. The isolation efficiency of the circuit, as a function of input drive power and frequency, is displayed in Figs.3(c),(d). This modification yields a twofold enhancement in the circuit's spectral bandwidth, expanding it from roughly 100 MHz to 200 MHz. Moreover, there is a minor enhancement in the power bandwidth, increasing by 1 dB. In comparison to the previous design, the novel circuit demonstrates a marginal increase in amplitude efficiency.

It is important to note that the anticipated improvement is attainable through a pragmatic alteration in the circuit, which can be readily implemented experimentally. Furthermore, the circuit provides flexibility by incorporating an additional capacitor, facilitating the adjustment of the resonant frequency via the tuning of the shunt capacitance. This shunt capacitor introduces an asymmetry in the coupling between the resonator and the transmission line, generating interference patterns that alter the resonant frequency response. As a result, the resonant frequency of Fano resonators deviates from the expected value for Lorentzian resonators possessing identical geometric parameters.

It is worth noting that this type of nonreciprocal element has been subjected to criticism in the literature for two reasons related to the origin of the nonreciprocity effect. Firstly, it requires the simultaneous absence of reflected waves or noise in the backward direction, which can be achieved through good matching, and a low level of temperature noise, which is ensured by installation in the dilution refrigerator[21]. Furthermore, external wideband noise can be filtered out by an additional resonator or Purcell filter, and the resonant response of the proposed isolator that acts as a bandpass filter. Secondly, for the same reason, this approach does not allow for the generalization of the isolator to a circulator with three or more ports in the CW regime of operation. However, it should be noted that our circulator is intended to operate in the pulsed regime, where the realization of nonlinear passive circulators is both possible and highly feasible[22].

In this work, we investigated quantum nonreciprocity using Lorentz and Fano circuits consisting of two qubits with intrinsic nonlinearity, frequency detuning, and coupling to a transmission line through capacitors. We validated the circuits' parameters by comparing them with existing experimental results and rigorous quantum simulations in QuCAT, confirming their close-to-reality behavior. By introducing a shunt capacitor, we transformed the Lorentz-type qubits circuit into a Fano-type circuit with an asymmetric spectral response, resulting in a significant enhancement in isolation up to 40 dB, twice the spectral bandwidth improvement, and a moderate increase in power bandwidth. This improvement in spectral bandwidth is especially crucial since conventional Faraday isolators have limited bandwidth, whereas microwave circuit QED operates in pulsed regimes. The enhanced power bandwidth could prove crucial for future research aimed at stabilizing nonlinear isolators in the presence of excitation noise and fluctuations. These results offer a promising avenue for the development of efficient passive quantum isolators.

**References**

[1] F. Arute *et al.*, "Quantum supremacy using a programmable superconducting processor," *Nature*, vol. 574, no. 7779, pp. 505–510, Oct. 2019, doi: 10.1038/s41586-019-1666-5.

[2] E. Gibney, "Hello quantum world! Google publishes landmark quantum supremacy claim," *Nature*, vol. 574, no. 7779, pp. 461–462, Oct. 2019, doi: 10.1038/d41586-019-03213-z.

[3] H. P. Paudel *et al.*, "Quantum Computing and Simulations for Energy Applications: Review and Perspective," *ACS Eng. Au*, vol. 2, no. 3, pp. 151–196, Jun. 2022, doi: 10.1021/acsengineeringau.1c00033.

[4] J. C. Bardin, D. H. Slichter, and D. J. Reilly, "Microwaves in Quantum Computing," *IEEE J. Microwaves*, vol. 1, no. 1, pp. 403–427, Jan. 2021, doi: 10.1109/JMW.2020.3034071.

[5] Y. Y. Gao, M. A. Rol, S. Touzard, and C. Wang, "Practical Guide for Building Superconducting Quantum Devices," *PRX Quantum*, vol. 2, no. 4, p. 040202, Nov. 2021, doi: 10.1103/PRXQuantum.2.040202.

[6] D. M. Pozar, *Microwave engineering, 4th Edition*. John Wiley & Sons, Inc., 2011.

[7] S. V. Kutsaev, A. Krasnok, S. N. Romanenko, A. Y. Smirnov, K. Taletski, and V. P. Yakovlev, "Up-And-Coming Advances in Optical and Microwave Nonreciprocity: From Classical to Quantum Realm," *Adv. Photonics Res.*, vol. 2, no. 3, p. 2000104, Mar. 2021, doi: 10.1002/adpr.202000104.

[8] A. Kord, D. L. Sounas, and A. Alu, "Microwave Nonreciprocity," *Proc. IEEE*, vol. 108, no. 10, pp. 1728–1758, Oct. 2020, doi: 10.1109/JPROC.2020.3006041.

[9] N. Nefedkin, M. Cotrufo, A. Krasnok, and A. Alù, "Dark-State Induced Quantum Nonreciprocity," *Adv. Quantum Technol.*, vol. 5, no. 3, p. 2100112, Mar. 2022, doi: 10.1002/qute.202100112.

[10] D. L. Sounas and A. Alù, "Non-reciprocal photonics based on time modulation," *Nat. Photonics*, vol. 11, no. 12, pp. 774–783, Dec. 2017, doi: 10.1038/s41566-017-0051-x.

[11] F. Fratini *et al.*, "Fabry-Perot Interferometer with Quantum Mirrors: Nonlinear Light Transport and Rectification," *Phys. Rev. Lett.*, vol. 113, no. 24, p. 243601, Dec. 2014, doi: 10.1103/PhysRevLett.113.243601.

[12] J. Dai, A. Roulet, H. N. Le, and V. Scarani, "Rectification of light in the quantum regime," *Phys. Rev. A*, vol. 92, no. 6, p. 063848, Dec. 2015, doi: 10.1103/PhysRevA.92.063848.

[13] A. Rosario Hamann *et al.*, "Nonreciprocity Realized with Quantum Nonlinearity," *Phys. Rev. Lett.*, vol. 121, no. 12, p. 123601, Sep. 2018, doi: 10.1103/PhysRevLett.121.123601.

[14] D. L. Sounas, J. Soric, and A. Alù, "Broadband passive isolators based on coupled nonlinear resonances," *Nat. Electron.*, vol. 1, no. 2, pp. 113–119, Feb. 2018, doi: 10.1038/s41928-018-0025-0.

[15] M. F. Gely and G. A. Steele, "QuCAT: quantum circuit analyzer tool in Python," *New J. Phys.*, vol. 22, no. 1, p. 013025, Jan. 2020, doi: 10.1088/1367-2630/ab60f6.

[16] B. D. Josephson, "Possible new effects in superconductive tunnelling," *Phys. Lett.*, vol. 1, no. 7, pp. 251–253, Jul. 1962, doi: 10.1016/0031-9163(62)91369-0.

[17] X. Gu, A. F. Kockum, A. Miranowicz, Y. Liu, and F. Nori, "Microwave photonics with superconducting quantum circuits," *Phys. Rep.*, vol. 718–719, pp. 1–102, Nov. 2017, doi: 10.1016/j.physrep.2017.10.002.

[18] A. Blais, A. L. Grimsmo, S. M. Girvin, and A. Wallraff, "Circuit quantum electrodynamics," *Rev. Mod. Phys.*, vol. 93, no. 2, p. 025005, May 2021, doi: 10.1103/RevModPhys.93.025005.

[19] P. Krantz, M. Kjaergaard, F. Yan, T. P. Orlando, S. Gustavsson, and W. D. Oliver, "A quantum engineer's guide to superconducting qubits," *Appl. Phys. Rev.*, vol. 6, no. 2, p. 021318, Jun. 2019, doi: 10.1063/1.5089550.

[20] J. Q. You and F. Nori, "Atomic physics and quantum optics using superconducting circuits," *Nature*, vol. 474, no. 7353, pp. 589–597, 2011, doi: 10.1038/nature10122.

[21] Y. Shi, Z. Yu, and S. Fan, "Limitations of nonlinear optical isolators due to dynamic reciprocity," *Nat. Photonics*, vol. 9, no. 6, pp. 388–392, Jun. 2015, doi: 10.1038/nphoton.2015.79.

[22] G. D'Aguanno, D. L. Sounas, H. M. Saied, and A. Alù, "Nonlinearity-based circulator," *Appl. Phys. Lett.*, vol. 114, no. 18, p. 181102, May 2019, doi: 10.1063/1.5094736.